# How Do Interactive Virtual Operas Shift Relationships between Music, Text and Image ?

**A. Bonardi**
Université Paris IV-Sorbonne
5, impasse du Débarcadère
78000 Versailles - France
alain.bonardi@wanadoo.fr

**F. Rousseaux**
Université de Reims
20, rue de Condé
75006 Paris - France
francis.rousseaux@univ-reims.fr

**Abstract**

In this paper we present the new genre of interactive operas implemented on personal computers. They differ from traditional ones not only because they are virtual, but mainly because they offer to composers and listeners new perspectives of combinations and interactions between music, text and visual aspects.

## 1    What Are Virtual Operas ?

We define as a virtual interactive opera any opera implemented on a personal computer, enabling some interactivity with its audience. This new genre is not based on the simple transposition of existing operas in the framework of new computing technologies; it takes into account the shiftings of uses and meanings induced by multimedia computing, assuming they could arouse new writings and ways of creativity [Bonardi 1998b].

Interactivity is based on free choices among different paths. Only open forms that have specially been designed for listeners can enable it. This specification leads us to give up the classical narrative "aristotelician" mode. This implies two possibilities of formal and temporal articulation :

- either consider the hypertext model and possibilities of automatic text generation [Balpe & al. 1996] to design an opera where hypermedia components strongly depend on text; this is the case in Jean-Pierre Balpe's *Barbe-Bleue* project,

- or take into account the specificities of computer-aided writing to design a new kind of "musical action" as Luciano Berio [Berio & Eco 1994] defined it : "Between a musical action and an opera, there are substantial differences. Opera is based on a 'aristotelician-like' narrative mode, which tends to have priority on musical development. On the contrary, in a musical action, the musical process rules the story". We have chosen this very direction for our research and our *Virtualis* project. In this interactive opera, the listener wanders in an open space, being essentially guided by metaphors of music rather than combinations of narratives.

In our paper, we deal with relationships between music, text and image, comparing what they used to be in former classical operas and what they could become in this new genre.

## 2 Relationships between Music, Text and Visual Aspects in Traditional Operas

Many authors have noticed that opera has been oscillating for four centuries between music predominance and text predominance. However, if we consider individual works, we can state that each opera has its own balance between music and text, and moreover that it remains constant during the whole work [Bonardi 1997]. It often has to do with the macroscopic form of operas, either based on musical constraints or on dramatic ones.

On the other hand, analysts like Michel Poizat [Poizat 1986] claim that the main dialectic in opera consists of tension between an homeostatic delight principle based on the intelligibility of language and a pleasure principle, as an asymptote of an exaltation drifting without any back force towards a sung shout. This does not contradict the former statement if we consider that the interaction between text and music in traditional operas as multiple, happening on various levels.

For instance, if we examine *The Magic Flute* by Mozart, we find that the general design of the work matches musical constraints, which is consistent with Mozart's position in his correspondence which claims that poetry should be the obedient daughter of music. In the precise case of *The Magic Flute*, let us notice that the predominance of music reinforces the philosophical aspects of the work since many philosophical symbols are encoded in the score more or less explicitly. Though singing only two arias, the Queen of the Night is a very interesting character since she actually embodies the conflict between delight and pleasure. In the second act, she encourages her daughter Pamina to murder Sarastro : this is musically emphasized by a progression at the beginning of the aria, accumulating an extreme tension. The surprise comes from the resolution of this tension, which does not lead to a climax, but to what Poizat calls the scansion of the singing exercise, with highly broken virtuosic passages. In fact, Mozart avoids the climax, which should be like a sung shout, but is not suitable at the classical age. Mozart goes further and nearly gives up words to the benefit of music in these virtuosic passages.

Concerning the importance of visual aspects in traditional operas, we could say that they are necessary but not sufficient. For instance, the audience often seems frustrated when an opera is played in a concert version. This means we need sets, costumes and lighting. But they just build a framework where music confronts text.

## 3 Opera and Interactivity

### 3.1 Collective Interaction

The first introduction of interactivity in opera happened in 1968, when *Votre Faust*, by Henri Pousseur and Michel Butor was created. In this "Fantaisie variable" written for five actors, four singers, twelve musicians and magnetic tape, some chances of interaction were given to the audience, so that the plot could be altered. One can really wonder whether selecting paths in an open work can be a collective action : can an opera be based on an interactive process driven by a kind of vote ? Is it possible to import a political paradigm, the democratic vote, as a principle of artistic design ?

### 3.2 Individual feedback

Contrary to this approach based on collective interaction giving average results, we claim that interactivity must also lead to individual feedback, that is, even though several people can take part together to the same process at the same time, each of them gets both collective and individual responses.

Using computers seems to be an interesting way to achieve this goal, in the framework of what we just named virtual interactive operas. Let us first notice that, with personal computers, the interaction necessarily happens through graphical elements. This means that visual aspects are predominant in these new operas, and therefore the main dialectic does not oppose music to text but images to another component, which can either be an hypertextual set or a musical set. As we said in the introduction of this paper, we are interested in the second kind of dialectic, between images and music.

If we want this interaction to happen in an open work, we have to give users possibilities of navigation among different music entities, that is creating a kind of "hypermusic" environment, which is quite different from an hypertextual one. Though music is a language based on a syntax [Boucourechliev 1993], it has no other meaning than itself. And yet, in the hypertextual environment, information nodes often contain textual pointers, able to designate or describe other information nodes in the user's context. These hypertextual links are obviously semantic. As a music entity is unable to designate by itself other entities, we need graphical metaphors to enable this navigation.

Text then takes a new role, which has less to do with narrative aspects than with sound and graphical ones. It means that we are closer to poetry than to theater. Indeed, words can be used in this context as sound resources or even as elements of the set.

## 4    Using Graphical Metaphors of Music

### 4.1    The Different Roles of Graphical Metaphors

As we just said, the role of graphical metaphors is to enable users to navigate through music. Offhand they have three different roles of mediation between music and listeners :

- the first role is to make explicit musical structures and processes handled by composers and analysts; metaphors are then used as translators.

- the second approach is interested in the listener's point of view : in a given context, what does the listener perceive, which categories are relevant for him ? That was Pachet's [Pachet & Delerue 1998] basic idea when developing their MIDI Spatializer named "Midispace" which enables users to move instruments on a stage, the program taking into account consistency constraints simulating the action of a sound engineer.

- the third role consists in arousing a kind of convergence between music and graphism, without making it necessarily explicit. The listener does not then know how to designate the musical phenomenon he has perceived, but he is able to associate it with geometric or color configurations.

Let us first say that these three aspects are often mixed. For instance, some researches in the field of psychoacoustics have shown that such structural elements as the tempo and the mode of a piece have an important influence on the emotion aroused, sometimes more than the musician's interpretation [Mc Adams & Bigand 1994].

### 4.2    Using GUIDO music notation language

Being interested in metaphors of structural elements of music, we address the problem of music annotation. We implemented an extension of the GUIDO music notation language [Hoos & al. 1998]. The main advantage of GUIDO is to enable the direct handling of aggregations, either temporal aggregations as phrases, or vertical aggregations as chords. Contrary to other descriptions [Mendes 1999], GUIDO does not separate elementary elements (for instance notes) from aggregations.

Let us consider for instance the very beginning of the *Diabelli Variations* for piano by Beethoven (opus 120).

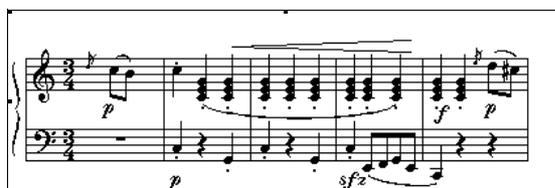

*figure 1 : beginning of the Diabelli Variations by Beethoven*

Here are the basic rules to use GUIDO :
- what is played sequentially is encoded with brackets : []
- what is played simultaneously is encoded with braces {}
- notes are encode using english notation (rests are represented by _)
- durations are indicated by the corresponding ratios
- octave and duration information are implicit : C1/4 D E indicates to play three quarters C, D, E in the same octave
- there are tags (for instance \staff) that enable to indicate various levels of information.

This extract by Beethoven should be encoded like this :
{[\staff<1> \slur<D2/16 C B1/8> C2/4 \slur<{C1 E G} {C E G} etc…>] [\staff<2> _/4 C0 _ G-1 C0 _ G-1 C0 \slur <E-1/8 F G E C/4>]}

### 4.3 Two Examples of Graphical Metaphors

We can build various graphical metaphors from GUIDO annotation files. First of all, we have thought of tunnels that would be a synthesis of many musical properties as shown below :

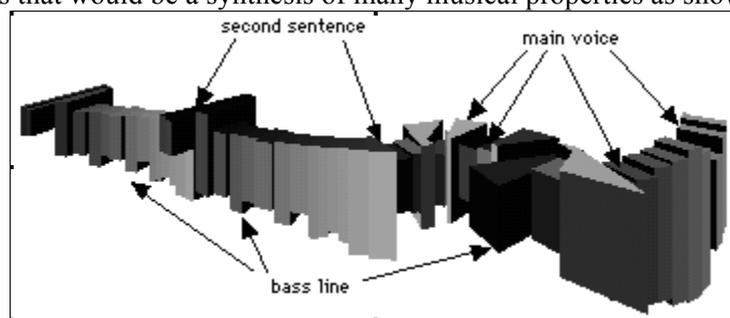

*figure 2 : tunnel built according to the beginning of Beethoven's piece*

In this case, we have the following mapping : the roof follows the main line, the floor follows the bass line, the bends and the colors correspond to the phrases, and the width to the vertical density.

We were also interested in visualizing polyphonies. Let us consider an extract of Mozart's *Piano Sonata K.282* (first part) :

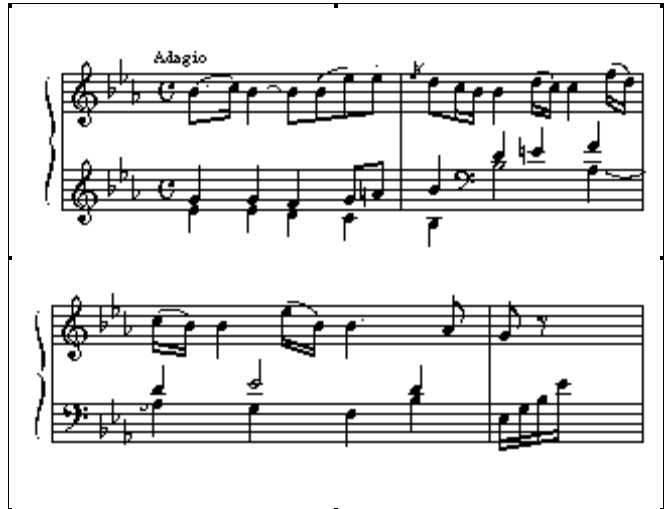
*figure 3 : extract of Mozart's Piano Sonata (first part)*

For this piece, we used the metaphor of roads or slides for each voice. The left hand part has been split into two voices, and the whole is displayed as a three voice piece. In the following example, three different instruments (piano, guitar, cello) have been attributed to the three voices and appear on the ground of each slide.

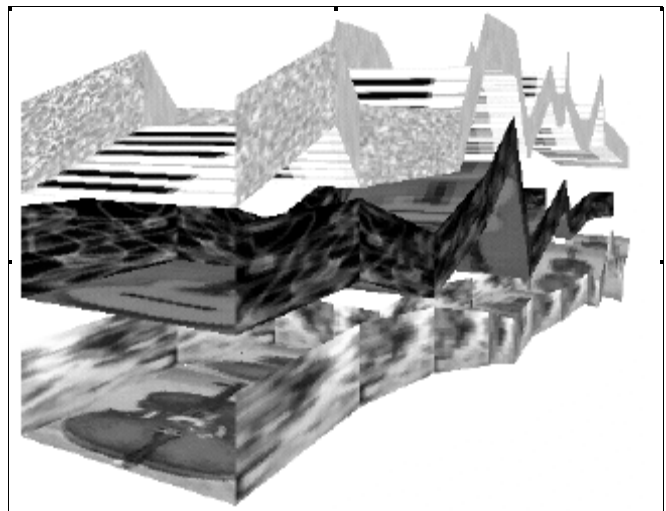
*figure 4 : vizualisation of Mozart's Piece*

These representations can also be used efficiently in the case of vocal music, and especially opera. Following the same principles, we have generated corridors that match each singing part. In this case, they also support semantic links, as pictures of the characters and some lyrics are displayed on the walls and are clickable. This enables the creation of a kind of « hyper-vocal music ». The two examples given come from Mozart's Magic Flute, more precisely from Tamino's Aria at the beginning of the first act.

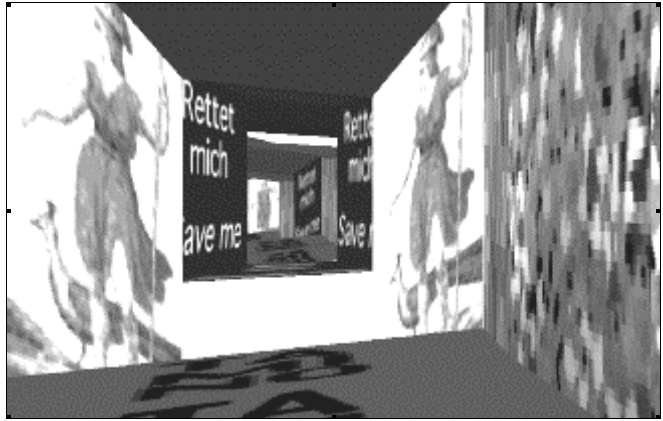

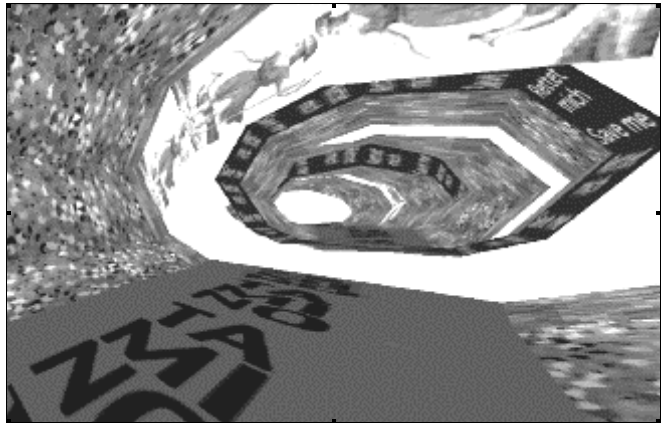

*figures 5, 6 : examples of vizualisation of Tamino's Aria in the first act*

## 6    The ALMA Environment

Handling music in computer-aided open forms requires a specific environment for creation. It first implies that music cannot be considered any longer as a resource stored in files, which is the common way it is processed in such multimedia authoring softwares as Director (Macromedia). Moreover, these authoring environments do not enable the composer to have a clear apprehension of then open form he wants to create since they were not designed for music purposes.

We have therefore started developing our own authoring system, named ALMA (as a tribute to Alma Mahler and also Gustav Mahler). This is a hierarchical object-oriented system where logical descriptions of music or stage indications (encoded thanks to an extension of the GUIDO language presented above) are separated from physical files to be played, either music files (audio or MIDI files) or graphics files (we can choose a metaphor for every piece, for the moment only tunnels or roads as shown above are available).

In this environment, text may have three different roles :
- either used for textual descriptions of music or stage indications using extended Guido,
- either used as a sound : for instance, \text("if music be the food of love") can be associated with different audio files from different speakers saying "if music be the food of love",
- or become a graphical element, thanks to graphical metaphors.

The ALMA system includes different modules :
- two annotation modules; the first one enables to add such descriptions as slurs or harmonic data to MIDI files; the second one is used to annotate audio files thanks to a corresponding MIDI file that was previously annotated with the first module.

- a COMPOSER module to edit and put together entities and links. Entities can be edited through four windows : a first one to edit texts written with the description language (extended GUIDO); a second window for the score; a third to watch the result of the graphical metaphor chosen for this very entity; a last one to manage physical resources (sound files, 3D generation modules, …). Let us notice that links can be either sequential (telling that one entity should be played after another), or based on a loop (so that an entity should be repeated several times) or conditional (with tests).
- a PLAYER module that performs the interactive score prepared with the COMPOSER module.

## Conclusion

In the context of new virtual operas, traditional dialectics and focuses are shifted by the predominance of images, and the new temporal dimension given by open forms. We are only beginning our exploration of new fields of interactive and multimedia musical composition, also experimenting new relationships between operas and their listeners.

## References

Auffret, Gwendal, Jean Carrive, Olivier Chevet, Thomas Dechilly, Rémi Ronfard & Bruno Bachimont (1999), *Audiovisual-based Hypermedia Authoring : using structured representations for efficient access to AV documents*, Proceedings of the 10th ACM Hypertext'99 Conference, 169-178

Balpe, Jean-Pierre, Alain Lelu, Frédéric Papy & Imad Saleh (1996), *Techniques avancées pour l'hypertexte*, Paris : Hermès

Berio, Luciano & Umberto Eco (1994), *Eco in ascolto - Entretien avec Luciano Berio*, in *Musique : texte, les cahiers de l'Ircam - Recherche et Musique n°6*, Paris : Ircam

Bonardi, Alain (1997), *Vers un opéra interactif*, Mémoire de DEA, formation doctorale "Musique et Musicologie du XX° siècle" EHESS/Paris IV Sorbonne

Bonardi, Alain & Francis Rousseaux, (1998), *Premiers pas vers un opéra interactif*, Proceedings of the « Journées d'Informatique Musicale 1998 » (JIM 98), LMA Publication n°148, Marseille, CNRS

Bonardi, Alain & Francis Rousseaux (1998), *Towards New Lyrical Forms*, in the Papers from the 1998 AAAI Fall Symposium - Technical Report FS-98-03, Menlo Park (USA, California) : AAAI

Boucourechliev, André (1993), *Le langage musical*, Paris : Fayard

Eco, Umberto (1962), *L'opera aperta*, Milan : Bompiani

Elliott, Conal (1999), *Modeling Interactive 3D and Multimedia Animation with an Embedded Language*, forthcoming in the *IEEE Transactions on Software Engineering* 1999, can be browsed at the following URL : http://www.research.microsoft.com/~conal/papers/ds19

Hoos, Holger, Keith Hamel, Kai Flade & Jurgen Killian (1998) *GUIDO Music Notation - Towards an Adequate Representation of Score Level Music*, Proceedings of the « Journées d'Informatique Musicale 1998 » (JIM 98), LMA Publication n°148, Marseille, CNRS

Mc Adams, Steve & Emmanuel Bigand (1994), *Penser les sons. Psychologie cognitive de l'audition*, Paris : Presses Universitaires de France

Mendes, David (1999), *Knowledge Representation Suitable for Music Analysis*, Proceedings of the « Journées d'Informatique Musicale 1999 » (JIM 99), Publication du CEMAMu, Issy-les-Moulineaux

Pachet, François & Olivier Delerue (1998), *A constraint-based Temporal Music Spatializer*, 10th ACM Multimedia Conference Proceedings

Poizat, Michel (1986), *L'opéra ou le cri de l'ange*, Paris : Editions A.M. Métailié

Pousseur, Henri (1997), *Musiques croisées*, Paris : L'Harmattan

Winkler, Tod (1998), *Composing Interactive Music, Techniques and Ideas Using Max*, Cambridge (Massachusetts, USA) : The MIT Press